\documentclass[12pt,notoc]{JHEP3}

\usepackage{amsmath,amssymb,euscript,array,cite}

\input{epsf}
\newcommand{\startappendix}{
\setcounter{section}{0}
\renewcommand{\thesection}{\Alph{section}}}
\newcommand{\Appendix}[1]{
\refstepcounter{section}
\begin{flushleft}
{\large\bf Appendix \thesection: #1}
\end{flushleft}}
\usepackage{epsfig}

\def\aD{{\dot\alpha}}

\def\N{{\cal N}}

\def\det{{\rm det}}

\def\Dbarslash{\,\,{\raise.15ex\hbox{/}\mkern-12mu {\bar D}}}
\def\Dslash{\,\,{\raise.15ex\hbox{/}\mkern-12mu D}}
\def\delslash{\,\,{\raise.15ex\hbox{/}\mkern-9mu \partial}}
\def\delbarslash{\,\,{\raise.15ex\hbox{/}\mkern-9mu {\bar\partial}}}

\newcommand{\MAT}[1]{\begin{pmatrix} #1\end{pmatrix}}
\newcommand{\EQ}[1]{\begin{equation} #1 \end{equation}}

\newcommand{\SP}[1]{\begin{equation}\begin{split} #1
\end{split}\end{equation}}

\title{Sitting on the Domain Walls of $\N=1$ Super Yang-Mills}
\author{Adi Armoni and Timothy J. Hollowood \\
Department of Physics,\\ University of Wales Swansea,\\
Swansea, SA2 8PP, UK.\\
E-mail: {\tt a.armoni,t.hollowood@swan.ac.uk}}
\preprint{SWAT/05/433}
\abstract{In pure $\N=1$ supersymmetric Yang-Mills with gauge group $SU(N)$, 
the domain walls which separate the $N$ vacua have been argued, on the
basis of string theory realizations, to be D-branes for the confining
string. In a certain limit, 
this means that a configuration of $k$ parallel domain walls is
described by a $2+1$-dimensional $U(k)$ gauge theory.
This theory has been identified by Acharya and Vafa as
the $U(k)$ gauge theory with 4 supercharges broken by a Chern-Simons
term of level $N$ in such a way that 2 supercharges are preserved. We
argue further that the gauge coupling of the domain wall gauge theory
goes like $g^2\sim\Lambda/N$, for large $N$.   
In the case of two domain walls, we show that the $U(2)$ world-volume
theory generates a quadratic potential on the Coulomb branch 
at two loops in 
perturbation theory which is consistent with there being a
supersymmetric bound state of the two wall system. A mass gap
of order $\Lambda/N$ is generated around the supersymmetric minimum
and we estimate the size of the bound-state to be order
$\Lambda/\sqrt{N}$.  
At large distance the potential reaches a constant that can
qualitatively account for the binding
energy of the two walls even though stringy effects are not, strictly
speaking, decoupled.}

\begin{document}

\section{Introduction}

The theory of D-branes in string theory has provided a remarkable
bridge between string theory and gauge theories. We now understand how
in certain situations the fundamental string {\em is\/} the confining
string of the gauge theory. In this context a D-brane will be object
on which the confining string can end. Such a set up was first 
described in \cite{Witten:1997ep}. Witten argued that type IIA D-branes
become the QCD D-brane. An explicit construction of the world-volume
theory on these QCD D-branes has been provided in \cite{Acharya:2001dz}.
In this case the gauge theory in the $3+1$ flat dimensions is 
pure $\N=1$ supersymmetric Yang-Mills. D4-branes wrapped on the
internal $S^2$ correspond to 2-brane in $3+1$ which are naturally
identified with the domain walls that separate the $N$ vacua of the
pure $\N=1$ theory. In fact such objects are known independently of
any string theory construction to be BPS objects preserving half the
supersymmetries \cite{Dvali:1996xe}. In brief, the vacua are labeled by the
value of the gluino condensate \cite{Shifman:1987ia,Davies:1999uw}
\EQ{
\langle\lambda\lambda\rangle_j=N\Lambda^3e^{2\pi ij/N}\ ,
}
$j=0,1,\ldots,N-1$. The basic domain wall separates the $j$-th
and $j+1$-th vacua. However, there are BPS bound states which
separate the $j$-th and $j+k$-th. The tension of the
bound state is determined by a central charge to be \cite{Dvali:1996xe}
\EQ{
T_k=\frac{N^2\Lambda^3}{4\pi^2}\sin\frac{\pi k}{N}\ .
\label{tk}
}
$k=1,\ldots,N-1$. Since the domain walls are D-branes, their
collective dynamics should---at least for small separations---be
described by the lightest modes of the confining string stretched
between them. This means that the configuration of $k$ basic parallel
walls should be described in terms of a $U(k)$ gauge theory in 
$2+1$ dimensions. Since the bulk gauge theory has four supercharges, the
domain wall theory should preserve half of these. The domain wall
theory was identified by Acharya and Vafa \cite{Acharya:2001dz} as the
$\N=2$ supersymmetric $U(k)$ gauge theory with a supersymmetrized
Chern-Simons term of level $N$ which only preserves half of the
supersymmetries. The Chern-Simons term arises from the $N$ units of RR
2-form flux through the $S^2$.

The purpose of this paper is to consider the potential energy between
two domain walls at a distance $X$ apart 
and in particular to understand the origin of the
binding energy of the two walls within the large $N$ expansion
\EQ{
\Delta T=2T_1-T_2=\frac{\pi\Lambda^3}{4N}\Big(1+{\cal O}(1/N^2)\Big)\ .
\label{be}
} 
In the string picture, the potential can be calculated by considering
closed string diagrams with boundaries on the two walls. The
topological expansion is basically an expansion in $1/N$ since 
we expect $g_s\sim1/N$. The leading contribution ${\cal
  O}(g_s^0)$ comes from the annulus digram illustrated in Figure 1.
\begin{figure}[ht]
\centerline{\includegraphics[width=2.5in]{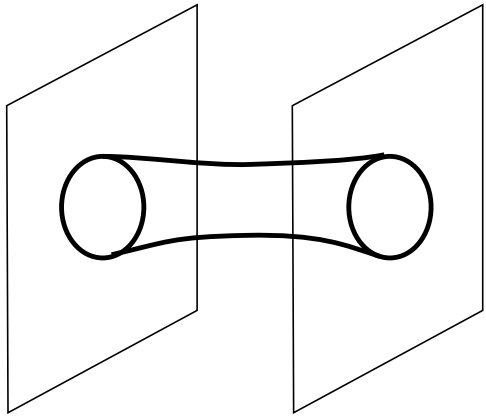}}
\caption{\footnotesize The annulus diagram.} \label{annulus}
\end{figure}
In more detail, the annulus diagram  
can be viewed either as closed string exchange or a
loop of open string. More precisely, there is
region of the moduli space where the closed string picture is
appropriate and a corresponding region where the open string picture
is valid. An explicit cut-off $\Lambda_0$, which is usually set around the 
string scale\footnote{The string scale is set by the tension of the
confining string $\sigma$ which is expected to be 
$\sim\Lambda^2$ at large $N$.} 
$\Lambda$ can be introduced which implements the
dichotomy of the diagram. The complete contribution is then a sum
\EQ{
V(X)=V^\text{closed}(X,\Lambda_0)+V^\text{open}(X,\Lambda_0)
\label{decom}
}  
which is independent of $\Lambda_0$.
We can, following \cite{Douglas:1996yp}, introduce some diagrammatic
notation to illustrate the partitioning of the 
moduli space of a string digram. In this notation
an open string is
represented by a double line where the Chan-Paton factors can take
one of two values corresponding to each of the two walls. 
A closed string is a wavy line. 
There are four vertices and their dependence on $g_s$ is 
illustrated in Figure 2.
\begin{figure}[ht]
\centerline{\includegraphics[height=5cm]{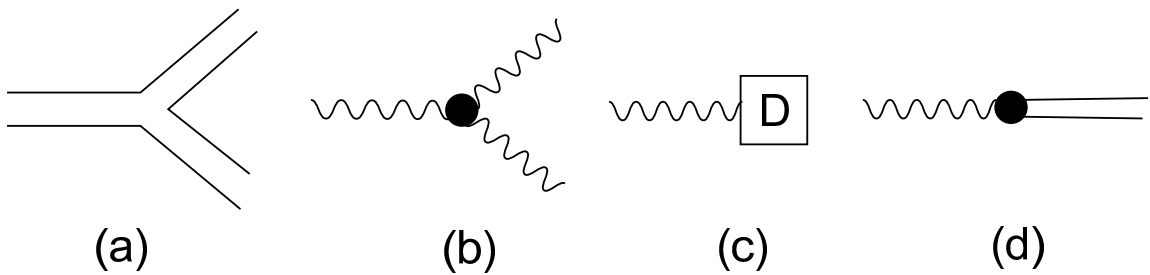}}
\caption{\footnotesize The open/closed string 
vertices. (a) $g_s^{1/2}$, (b) $g_s$, (c)
$g_s^0$, (d) $g_s^{1/2}$.} \label{vertices}
\end{figure}
The annulus diagram splits into two open/closed string graphs
realizing the dichotomy \eqref{decom} as illustrated in Figure 3.
\begin{figure}[ht]
\centerline{\includegraphics[height=3.5cm]{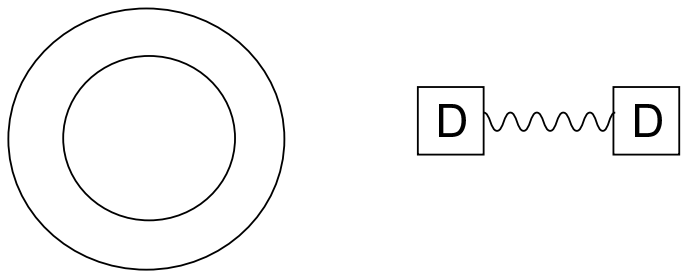}}
\caption{\footnotesize The dichotomy of the annulus into open/closed string 
graphs.} \label{annulus2}
\end{figure}
The effect of the cut-off $\Lambda_0$ has the effect of suppressing
contributions from higher mass open string modes on the potential 
so that the leading
contribution in the open string sector come from the lightest
modes. These modes are captured by the D-brane world-volume gauge
theory which in this case is the Acharya-Vafa gauge theory. As the
separation of the walls increases, the masses of open strings stretched
between the walls increases beyond $\Lambda_0$. This means that the
contribution to the potential from the open string graphs is suppressed at
large $X$. Actually we shall find that the one-loop contribution to
the effective potential of the Acharya-Vafa theory on its Coulomb branch
vanishes because of supersymmetry.
Consider now the closed string contribution. The 
closed strings correspond to glueballs in the gauge theory which are
massive since the gauge theory is expected to have a mass gap.
This means that there is a long range contribution to the
potential from glueball (closed string) exchange of the form
\EQ{
V^\text{closed}(X)\sim \sum_i\lambda_ie^{-M_iX}\ .
}
The effect of the cut-off $\Lambda_0$ is to soften the behaviour for
$X<\Lambda_0^{-1}$. 
However, as explained in  \cite{Armoni:2003jk},  glueballs 
always occur in supersymmetric multiplets appropriate to the
bulk theory: in this case the $\N=1$ theory in four dimensions. 
In other words, in chiral multiplets implying that 
glueballs always occur in degenerate even and odd parity
pairs. The exchange of even and odd parity
glueballs precisely cancels since the walls are BPS. This precisely
why the force between parallel D-branes in Type II string theory vanishes.
For example, the 
lightest glueball pair are expected to be scalars. The 
even parity partner couples to the wall tension $T$ and the
odd parity partner couples to the wall charge $Q$. Since the walls
are BPS they have $T=|Q|$. In our configuration
of two walls, the one on the left has tension/charge $(T,-Q)$ while
that on the right $(T,Q)$. So the potential for exchange of the lightest
even/odd parity pair glueballs of mass $M$ is 
\EQ{
V^\text{closed}(X)
\thicksim T^2e^{-M X}+Q(-Q)e^{-M X}=(T^2-|Q|^2)e^{-M X}=0\ .
}
This same reasoning extends to the whole
tower of glueballs 
and so the contribution of closed strings to the annulus vanishes.
To summarize, we expect the complete annulus contribution to the
potential to vanish and this 
is consistent with the binding energy \eqref{be} which
vanishes at ${\cal O}(N^0)$.

\begin{figure}[ht]
\centerline{\includegraphics[height=5cm]{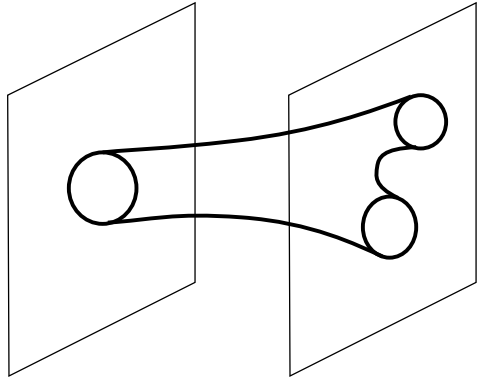}}
\caption{\footnotesize The pants diagram.} \label{pants1}
\end{figure}

\begin{figure}[ht]
\centerline{\includegraphics[height=8cm]{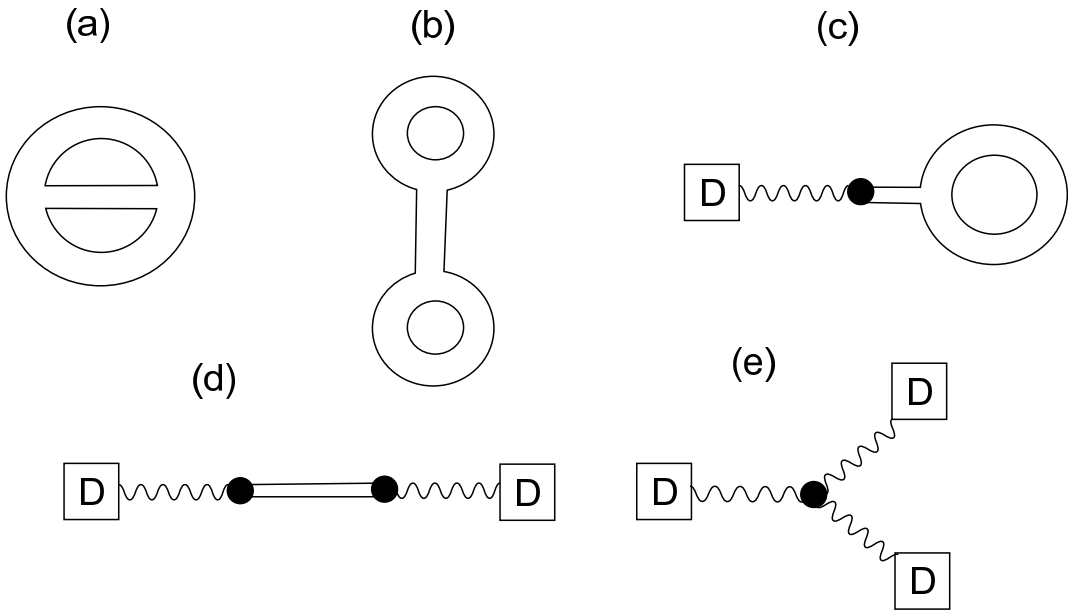}}
\caption{\footnotesize The partitioning of the pants diagram into
  open/closed string graphs.} \label{pants2}
\end{figure}

The next order in the $1/N$ expansion equates to
${\cal O}(g_s)$ corresponding to the ``pants diagram''
illustrated in Figure \eqref{pants1}.
In this case, there are five associated 
open/closed string graphs as illustrated
in Figure \eqref{pants2} (essentially a copy of 
Figure 4 of \cite{Douglas:1996yp}).
The leading contribution to the open
string graphs (a) and (b) corresponds to the two-loop effective potential of
the Acharya-Vafa theory on its Coulomb branch. The main result of this
paper is that this perturbative contribution to two loops is
\EQ{
V_\text{2-loops}(X)=\frac{2\pi c^6\Lambda^5 X^2}
{N(4\pi^2c^2+\Lambda^2X^2)}\ , \label{result}
} 
where $c$ is determined by the ratio $\sigma/\Lambda^2$ ($\sigma$ is
the tension of the confining string). This result has precisely
the right behaviour to account for the existence of a BPS bound-state
(it vanishes for $X=0$) and for the binding energy (the finite limit
as $X\to0$). The question is over what regime 
we can trust \eqref{result}? The situation is delicate because 
some of the fields in the Acharya-Vafa theory have a ``topological
mass'' of order the string scale and so there is no
genuine hierarchy between these modes and other excited open string modes.
Furthermore, the Acharya-Vafa theory is finite and \eqref{result} does
not need to be regulated. Since there is an explicit cut-off
$\Lambda_0$ there will be finite cut-off corrections as well as
corrections from higher derivative 
terms in the domain wall DBI action. We expect to be
able to trust it for small $X\ll\Lambda$ where the stringy cut-off
effects can be neglected. Notice that \eqref{result} has a minimum at
zero which is consistent with the expectation that two walls form a
BPS bound state. In addition, around the minimum the theory the
potential has the form 
\EQ{
V_\text{2-loops}(X)=\frac{c^4\Lambda^5 X^2}
{2\pi N}+\cdots \label{sma}
} 
and so the theory has a mass gap $\sim\Lambda/ N$. But what
about large $X$ where we wish to discover the binding energy? 
In this limit, the excited open string modes of mass
$M\sim \Lambda^2 X$ are
suppressed by $\exp(-M^2/\Lambda_0^2)$ \cite{Douglas:1996yp} but this
leaves open the question of cut-off and DBI corrections.
We have no definitive argument for why these effects could not alter the
asymptotic behaviour of \eqref{result} for large $X$:
\EQ{
V_\text{2-loops}(X)=\frac{2\pi c^6\Lambda^3}N-\frac{8\pi^3
  c^8\Lambda}{NX^2}+\cdots
}
The most
conservative interpretation is that \eqref{result} 
captures the qualitative behaviour of the potential from the open
string diagrams (a) and (b). What is clear is that these terms can
lead to the constant binding energy even though one normally expects
that the open string diagrams are suppressed at large $X$. In
particular, since there are no massless closed string modes we do not
expect the power law behaviour $1/X^{2p}$ in the subleading terms to 
survive stringy corrections. 

Now we turn to the two other sub-graphs (c), (d) and (e) in Figure
\eqref{pants2}. We expect (c) to vanish using the same argument we
applied to the annulus: glueballs come in even/odd parity pairs. 
Contributions $(d)$ and $(e)$, 
on the other hand, involve an interaction of the
glueballs with open string modes. In principle these interactions
could be investigated by couplings the Acharya-Vafa theory to the
appropriate closed string modes. What is incontrovertible is that the
Acharya-Vafa theory breaks the symmetry between even and odd parity glueballs
since the supersymmetry is broken by half on the walls. 
Consequently, as described in \cite{Armoni:2003jk}, there can be a
non-vanishing contribution from glueballs which would have the form 
\EQ{
V^{(d)+(e)}(X)\thicksim \frac1N\sum_i\lambda_ie^{-M_iX}
} 
for large $X$. 

Using the effective theory, we can estimate the size of the bound
state following \cite{Douglas:1996yp}. The thickness of a single
wall was estimated in refs.\cite{Dvali:1999pk,Gabadadze:1999pp}
as $\sim \Lambda / N$.
Ignoring factors of order one, near the supersymmetric minimum
\EQ{
S_\text{eff}\sim N\Lambda^3\int d^3x\,\Big(-(\partial_i
X)^2-\frac{\Lambda^2}{N^2}X^2+\cdots\Big)\ .
\label{eff}
}
The squared size of the bound state is related to the two-point function
of $X$. At leading order in $1/N$ we can ignore the mass term and all
the higher interactions in \eqref{eff}. At this order $X$ is a
massless free field, so
\EQ{
\langle X(\epsilon)X(0)\rangle \thicksim \frac1{N\Lambda^3\epsilon}
}
This is obviously singular as $\epsilon\to0$ so in order to make
sense of this result as the square of the size of the bound state, we
must cut it off at the only natural UV scale in the problem: the 
string scale $\epsilon\sim\Lambda^{-1}$.
The size of the bound-state at leading order in $1/N$ 
is then $N^{-1/2}\Lambda^{-1}$.

It is interesting to compare the domain wall story above with the discussion in
\cite{Douglas:1996yp} of the bound state of two D-strings and one
fundamental string. In this case the $SU(2)$ theory on the two D-branes
generates a potential with similar characteristics at the one-loop
level. There is a supersymmetric minimum with a mass gap 
and the potential then rises
to a constant. In this case, the one-loop result accounts for all
the binding energy of the system at leading order in $g_s$. One
difference is that the power-law behaviour at large $X$ matches the
effects of massless closed strings which are absent for the domain
wall case.

The paper is organized as follows: in Section 2 we introduce the
Acharya-Vafa theory. In section 3 we calculate the Coleman-Weinberg 
effective potential for the Acharya-Vafa theory on its Coulomb branch
up to two loops in perturbation theory.

\section{Domain Walls as D-branes}

In the Type IIA string theory construction of \cite{Acharya:2001dz}, 
the fundamental
string is the confining string of the gauge theory. The domain walls
correspond to D4-branes wrapped on an $S^2$ in the internal space;
hence they appear as 2-branes in the $3+1$-spacetime of the gauge
theory on which the confining string can end. Of course it is rather
non-trivial---nay mysterious---that a confining string can end on a
domain wall since ordinarily the confining string can only end on
fundamental colour charges. In our theory there are only adjoint-valued fields.
A heuristic mechanism was described by Witten in 
\cite{Witten:1997ep} (attributed originally to S.-J.~Rey). 
One imagines that the vacua are described by the
condensation of the rather elusive QCD monopole, or, for more general
vacua, dyon. The basic domain wall
separates vacua which differ by the transformation 
$\theta\to\theta+2\pi$. In the abelian case, such a transformation
shifts the electric charge of the monopole turning it into a 
dyon. Generalizing this to the non-abelian case
one suspects that 
the corresponding shift in the colour charge is equivalent to that of a
quark. A quark excitation can then appear in the domain
wall corresponding to a bound-state of a dyon, on one side, and an
anti-dyon, on the other. In this way, it is possible for a confining
string to end on a domain wall. In another related model, a domain wall 
can be considered as being composed of a net of baryon vertices, connected
by QCD-strings \cite{Armoni:2003ji}. 
This model provides an explanation for the $N$ dependence 
of the domain wall tension, as well as why a string can end on the wall.

We will take as a working hypothesis, the idea that the domain walls
are D-branes for the confining string. In this view,
the collective dynamics of the domain walls is described in the
conventional D-brane-like way 
by the lightest modes of the confining
string which end on the branes. For a configuration of $k$ domain walls
the resulting collective dynamics is a Born-Infeld type theory with a
$U(k)$ gauge field $A_i$, a single real adjoint-valued scalar field $X$ which
describes the transverse positions of the 2-branes in three-dimensional
space. There are also fermions and a particular Chern-Simons term
which we describe in detail below.

For the bosonic fields, the domain wall action will be of the
form\footnote{The following expression involves the symmetrized trace,
  which for our purposes can be replaced by an ordinary trace in the
  low-energy limit.}
\EQ{
S=\frac{T_k}k\int d^3\xi
\,\text{STr}\,\sqrt{-\det\big(\eta_{ij}+D_iXD_jX+
\sigma^{-1}F_{ij}\big)}+\cdots\ ,
} 
where the ellipsis represents fermionic terms and also---most
importantly---a Chern-Simons term. In the above, $\sigma$ is the tension of
the confining string. The normalization of the action has been fixed
by requiring that when the walls are in the ground state the tension
is $T_k$ as in \eqref{tk}. In principle, this pre-factor should be
determined by the way the D4 branes are wrapped on the internal space.
At low energies we can expand the square root in the
fluctuations to quadratic order:
\EQ{
S=T_kV_3+\frac{T_k}{k\sigma^2}\int
d^3\xi\,\text{Tr}\big(-\tfrac12(D_i\phi)^2-\tfrac14(F_{ij})^2+\cdots\big)\
,}
where $\phi=\sigma X$ is a real scalar of mass dimension 1. The
constant term proportional to the world volume $V_3$
is just the contribution from the tension of the walls in the ground
state. The effective gauge coupling 
of the theory is evidently
\EQ{
g^2=\frac{k\sigma^2}{T_k}\ .
}
For large $N$ (and fixed $k$), we have $\sigma=c\Lambda^2$ (for a numerical constant
$c$) while $T_k=kN\Lambda^3/(4\pi)$, and so
\EQ{
g^2\ \underset{N\to\infty}=\ \frac{4\pi c^2\Lambda}N\ .
\label{newrelation}
}
Notice that in the string theory interpretation $g^2\sim
\sigma^{1/2}g_s$. In other words, the string coupling
$g_s\sim 1/N$ as one expects.

We now turn our attention to the fermionic sector. Generically one would expect
the D4-branes wrapped on $S^2$ in the Type IIA set-up 
to preserve four supercharges on their
world-volume. It will ultimately turn out that this is not the case, but
for the moment suppose it were true. Then the theory on the
2-branes would then be the $\N=2$ supersymmetric 
gauge theory in $2+1$ dimensions. This can
be considered as the dimensional reduction of an $\N=1$ supersymmetric
$U(k)$ gauge
theory from four dimensions to three dimensions. We use Wess and Bagger
conventions in four dimensions and then dimensionally reduce by
removing dependence on $x^2$; so $\xi^i=(x^0,x^1,x^3)$. The component
$A^2$ is identified with the scalar $\phi$. The complex Weyl
fermion $\lambda_\alpha$ can be split into a real and imaginary part
in three dimensions:
\EQ{
\lambda_\alpha=\frac1{\sqrt2}(\chi_\alpha+i\psi_\alpha)\ ,\qquad
\bar\lambda_\aD=\frac1{\sqrt2}(\chi_\alpha-i\psi_\alpha)\ .
}
Note that in three dimensions one doesn't distinguish between a dotted
and un-dotted spinor index. The Lagrangian of the $\N=2$ theory is
then\footnote{To be clear, $\chi
  \Dslash\chi=\chi^\alpha\hat\sigma^i_{\alpha\beta}D_i\chi^\beta$ where
  $\hat\sigma^i=(\sigma^0,\sigma^1,\sigma^3)$ .}
\EQ{
{\cal
  L}_{\N=2}=\frac1{2g^2}\text{Tr}\big(-(D_i\phi)^2-\tfrac12(F_{ij})^2-i
\chi\Dslash\chi-i\psi\Dslash\psi-2\chi[\phi,\psi]\big)\ .
}
The key observation of Acharya and Vafa \cite{Acharya:2001dz} 
is that since there are $N$ units of RR 2-form flux through
the internal $S^2$ on which the D4-branes are wrapped, a Chern-Simons
term is induced in the 2-brane world-volume dynamics which breaks the
supersymmetry by half. In more detail, for the case of a single domain
wall, there is an interaction of the form
\EQ{
\int \tilde A\wedge F\wedge F=\int \tilde F\wedge A\wedge dA
}
on the $D4$-brane world-volume, where $\tilde A$ is the bulk RR gauge
potential and $F$ is the $U(1)$ field strength on the brane. Since
there are $N$ units of RR flux through the $S^2$, $\int_{S^2}\tilde F=N$,
and so there is an induced Chern-Simons term of level $N$ in the effective
theory in 3-dimensions. Extending this argument to many walls and
including the fermions, the full action proposed by Acharya and Vafa
is 
\SP{
{\cal
  L}_\text{AV}&=\frac1{2g^2}\text{Tr}\Big(-(D_i\phi)^2-\tfrac12(F_{ij})^2-i
\chi\Dslash\chi-i\psi\Dslash\psi-2\lambda[\phi,\psi]\\
&\qquad\qquad+
N\big(\epsilon_{ijk}(A^i\partial^jA^k+\tfrac13A^iA^jA^k)
+i\chi\chi\big)\Big)\ .
\label{av}
}
The multiplets of $\N=1$ supersymmetry are then $(A_i,\chi)$ and
$(\phi,\psi)$ and two supercharges that survive are obtained from the
four supercharges in four dimensions by taking the Grassmann parameter
of the supersymmetry transformation to be real. 

Classically at least, the theory on the domain walls has a Coulomb
branch on which the scalar field $\phi$ develops a VEV involving a
scale that we denote $\varphi$. If the
Chern-Simons term in \eqref{av} were absent then the theory would have
$\N=2$ supersymmetry and a homomorphic description. The real scalars
in the unbroken $U(1)^k$ gauge group naturally combine with the dual
photons to form $k$ complex scalars. In this case, any effective 
potential on the
Coulomb branch is determined by a superpotential holomorphic in the
complex scalars. Holomorphy then forbids the
generation of a potential on the Coulomb branch in perturbation
theory. However, a potential is generated by non-perturbative
instanton effects giving the classic runaway behaviour first uncovered
by Affleck, Harvey and Witten \cite{Affleck:1982as}. 
In the present theory, since we
only have $\N=1$ supersymmetry there are no holomorphic indulgences
and perturbative contributions to the effective
potential are not ruled out. 
There will also be non-perturbative instanton contributions,
however, since both $g^{-2}$ and the Chern-Simons coupling scale as $N$,
these will be exponentially suppressed at large $N$. In
what follows we shall be working for the most part 
at large $N$ 
and so we shall not discuss these non-perturbative contributions
any further. At finite $N$, the non-perturbative contribution would be
important for small values of the VEVs (short distances).

With this restriction in mind, our goal, therefore, is to
investigate the perturbative contribution to the effective potential
on Coulomb branch obtained by integrating out all the massive modes of
the Acharya-Vafa theory.
We shall only consider the case of two domain walls, in which case there
is a single VEV $\varphi$ which gives the separation between the domain
walls as $\varphi/\sigma$. 

We can already see the connection between the binding energy and
perturbation theory by expressing the former in terms of the natural
perturbative couplings of the Acharya-Vafa theory. 
These are the gauge coupling $g^2$, the
``topological mass'' 
\EQ{
m=g^2N=4\pi c^2\Lambda\Big(1+
\frac{a_1}{N^{2}}+\frac{a_2}{N^4}+\cdots\Big) 
}
as well as the Higgs mass $\varphi$.
The binding energy \eqref{be} written in terms of these parameters is
\EQ{
\Delta
T=\frac{\pi\Lambda^3}{4N}\Big(1+\frac{b_1}{N^2}+\frac{b_2}{N^4}+
\cdots\Big)\\
=\frac{g^2m^2}{2^8c^6}\Big(1+\frac{b_1g^4}{m^2}+\frac{b_2g^8}{m^4}+\cdots
\Big)\ .
\label{bee}
}
From this we can identify the leading effect at order $1/N$ 
as coming from two loops in
perturbation theory while the sub-leading terms of order
$1/N^{2i+1}$ as coming from $2(i+1)$ loops.

\section{The Two Wall Potential in Perturbation Theory}

In this section, we consider the 2-loop calculation of the effective
potential for the $U(2)$ Acharya-Vafa theory. The overall $U(1)$
factor is decoupled and for the purposes of the following calculation
we can consider the $SU(2)$ theory instead. On the Coulomb branch the scalar
field $\phi$ develops a VEV which breaks the gauge symmetry from
$SU(2)$ to $U(1)$. Using the basis $\{\tfrac12\tau^a\}$ 
for $SU(2)$,\footnote{We will also use the notation $\phi^\pm=\phi^1\pm
  i\phi^2$ for the charged components of a field.} we will choose a VEV
\EQ{
\phi=\phi^a\frac{\tau^a}2=\varphi\frac{\tau^3}2 .
}
After the Higgs mechanism and in the presence of the Chern-Simons term,
$\phi^3$ and $\psi^3$ remain massless. 
The other fields are either massive or the would-be Goldstone
bosons. We discuss them seriatim:

{\bf (1) Gauge bosons.} The charged components $A^\pm_i$ 
of the gauge bosons have a
complicated propagator which reflects a mixture between the Higgs
effect and the topological mass \cite{Deser:1981wh} 
arising from the Chern-Simons
term.\footnote{The Chern-Simons term in three dimensions gives gauge
  bosons a mass, the ``topological mass''  \cite{Deser:1981wh}. The
  situation with both a Chern-Simons term and spontaneous symmetry
  breaking has been much discussed in the literature: see, for
  example, \cite{Chen:1994zx,Chen:1995gk} and references therein.} 
In Euclidean space, which we now use throughout, and Landau
gauge, the propagator is\footnote{The propagator is diagonal in colour
  indices.} 
\EQ{
\Delta_{ij}^\pm(p)=\frac{(\delta_{ij}-p_ip_j/p^2)(p^2+\varphi^2)-
m\epsilon_{ijk}p_k}{(p^2+m_+^2)
(p^2+m_-^2)}\ ,
}
where 
\EQ{
m_\pm=\sqrt{\varphi^2+m^2/4}\pm m/2\ .
}
The neutral component $A^3_i$ has a purely topological mass and the
propagator in Landau gauge is
\EQ{
\Delta_{ij}^3(p)=\frac{(p^2\delta_{ij}-p_ip_j)-
m\epsilon_{ijk}p_k}{p^2(p^2+m^2)}\ .
}

{\bf (2) Scalars.} The neutral component $\phi^3$ is the massless Higgs
 field while $\phi^\pm$ are the would-be Goldstone Bosons and so 
are massless in Landau gauge.

{\bf (3) Fermions.} $\psi^3$ is massless while $\chi^3$ has mass $m$ arising
    from the supersymmetrized Chern-Simons term in \eqref{av}. 
The charged fermions $\psi^\pm$ and $\chi^\pm$ 
are mixed via the Yukawa coupling with the Higgs VEV. This creates the
    mass term
\EQ{
\MAT{\chi^- & \psi^-}\MAT{ m & -i\varphi \\ i\varphi & 0} 
\MAT{\chi^+\\ \psi^+}\ .
}
The eigenstates $\tilde\chi$ and $\tilde\psi$ of mass $m_+$ and $-m_-$
    are related to the original basis via
\EQ{
\MAT{\chi\\ \psi}^\pm=\sqrt{\frac2m}
\MAT{-m_+^{1/2} & m_-^{1/2}\\ 
i\varphi m_+^{-1/2} &
    i\varphi m_-^{-1/2}} 
\MAT{\tilde\chi\\ \tilde \psi}^\pm\ .
}

In addition to these fields and their interactions, we have to add the
usual gauge fixing terms and associated ghosts $(\bar c,c)$. The
vertices are those of a conventional spontaneously broken gauge theory
except that the
Chern-Simons term \eqref{av} modifies the momentum dependence of the 
three gauge vertex to
\EQ{
(p_1-p_2)_k\delta_{ij}+(p_2-p_3)_i\delta_{jk}+(p_3-p_1)_j\delta_{ik}
-m\epsilon_{ijk}
}
in Euclidean space.\footnote{In Euclidean space the
Chern-Simons term is pure imaginary.}

The effective potential as a function of the VEV $\varphi$ (which becomes the
field of the low-energy effective action) is obtained by integrating 
out all the massive modes: that is every field except $\phi^3$ and
$\psi^3$. In perturbation theory, the contribution is given by summing
all the vacuum graphs with massive fields propagating in the loops. It is
straightforward to verify that the one-loop contribution vanishes
identically due to the mass degeneracies entailed by
supersymmetry.  
At the two loop level, there are two kinds of vacuum graph; namely, the sunset
and the figure-of-eight illustrated in Figure \eqref{twoloops} below.
\begin{figure}[ht]
\centerline{\includegraphics[width=2.5in]{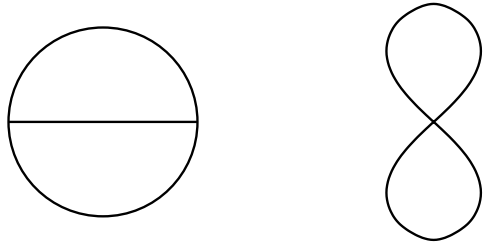}}
\caption{\footnotesize Vacuum graphs at two loops. } \label{twoloops}
\end{figure}
Although the theory
is finite, each separate graph is divergent and must be
regularized. Since we wish to preserve supersymmetry we use the dimensional
reduction regularization scheme.\footnote{In a nutshell, loop momenta
  propagate in $d$ dimensions, while the tensor and spinor structure
  is appropriate to 3 dimensions.} If the
diagrams are calculated correctly, the poles in $d-3$ cancel
to leave a finite result. Below we record the result for each separate
graph. The singular part of each graph appears in the combination
\EQ{
\log\hat\mu\equiv\frac12\Big(\frac1{3-d}
+1-\gamma+\log4\pi\Big)+\log\mu\ ,
}
where $\mu$ is the usual dimensionful parameter of dimensional regularization.
 
First of all, we have the following sunset graphs:\newline
{\bf (1)} $\phi^\pm$ - $\chi^3$ - $\tilde\chi^\pm$
\SP{
&\frac{g^2(\sqrt{m^2+4\varphi^2}-m)}{32\pi^2\sqrt{m^2+4\varphi^2}}
\Big(m^2+m\sqrt{m^2+4\varphi^2}
\\ &+(5m^2+2\varphi^2+3m\sqrt{m^2+4\varphi^2})
\log\frac{2{\hat\mu}}{3m+\sqrt{m^2+4\varphi^2}}\Big)\ .
}
{\bf (2)}  $\phi^\pm$ - $\chi^3$ - $\tilde\psi^\pm$
\SP{
&-\frac{g^2(\sqrt{m^2+4\varphi^2}+m)}{32\pi^2\sqrt{m^2+4\varphi^2}}
\Big(m^2-m\sqrt{m^2+4\varphi^2}
\\ &+(-5m^2-2\varphi^2+3m\sqrt{m^2+4\varphi^2})
\log\frac{2{\hat\mu}}{m+\sqrt{m^2+4\varphi^2}}\Big)\ .
}
{\bf (3)} $A^3$ - $\tilde\chi^\pm$ - $\tilde\chi^\pm$
\EQ{
-\frac{g^2}{32\pi^2}\Big(m^2-2\varphi^2+m\sqrt{m^2+4\varphi^2}+2(5m^2+4\varphi^2
+4m\sqrt{m^2+4\varphi^2})\log\frac{{\hat\mu}}{2m+\sqrt{m^2+4\varphi^2}}\Big)\ .
}
{\bf (4)} $A^3$ - $\tilde\psi^\pm$ - $\tilde\psi^\pm$
\EQ{
-\frac{g^2}{32\pi^2}\Big(-3m^2-2\varphi^2+3m\sqrt{m^2+4\varphi^2}+2(5m^2+4\varphi^2
-4m\sqrt{m^2+4\varphi^2})\log\frac{{\hat\mu}}{\sqrt{m^2+4\varphi^2}}\Big)\ .
}
{\bf (5)} $A^\pm$ - $\chi^3$ - $\tilde\chi^\pm$
\SP{
&-\frac{g^2}{16\pi^2(m^2+4\varphi^2)}\Big(m^4+6m^2\varphi^2+m(m^2-2\varphi^2)
\sqrt{m^2+4\varphi^2}\\ 
&+2m(4m^3+16m\varphi^2+(5m^2+4\varphi^2)\sqrt{m^2+4\varphi^2})
\log\frac{\hat\mu}{2m+\sqrt{m^2+4\varphi^2}}\\
&+(m^4+5m^2\varphi^2+4\varphi^4+(5m\varphi^2-m^3)
\sqrt{m^2+4\varphi^2})\log\frac{2{\hat\mu}}{3m+\sqrt{m^2+4\varphi^2}}\Big)\ .
}
{\bf (6)} $A^\pm$ - $\chi^3$ - $\tilde\psi^\pm$
\SP{
&-\frac{g^2}{16\pi^2(m^2+4\varphi^2)}\Big(m^4-10m^2\varphi^2+m(-m^2+6\varphi^2)
\sqrt{m^2+4\varphi^2}\\ 
&-2m(-4m^3-16m\varphi^2+(5m^2+4\varphi^2)\sqrt{m^2+4\varphi^2})
\log\frac{\hat\mu}{\sqrt{m^2+4\varphi^2}}\\
&+(m^4+5m^2\varphi^2+4\varphi^4-(5m\varphi^2-m^3)
\sqrt{m^2+4\varphi^2})\log\frac{2{\hat\mu}}{m+\sqrt{m^2+4\varphi^2}}\Big)\ .
}
{\bf (7)} $A^3$ - $\phi^\pm$ - $\phi^\pm$
\EQ{
-\frac{g^2m^2}{16\pi^2}\log\frac{\hat\mu} m\ .
}
{\bf (8)} $A^3$ - $A^\pm$ - $\phi^\pm$
\SP{
&-\frac{g^2}{64\pi^2m^2\sqrt{m^2+4\varphi^2}}\Big(14m^3\varphi^2+4m\varphi^4
-2m^4\sqrt{m^2+4\varphi^2}\log\frac{\hat\mu}
m\\ &+\varphi^2(5m^3+11m\varphi^2-(5m^2+\varphi^2)
\sqrt{m^2+4\varphi^2})\log\frac{\hat\mu}{-2m+\sqrt{m^2+4\varphi^2}}\\
&+(m^5+24m^3\varphi^2-22m
\varphi^4+(m^4+6m^2\varphi^2)\sqrt{m^2+4\varphi^2})\log
\frac{\hat\mu}{m+\sqrt{m^2+4\varphi^2}}\\ &+(-m^5+19m^3\varphi^2+11m\varphi^4+
(m^4+11m^2\varphi^2+\varphi^4)\sqrt{m^2+4\varphi^2})\log\frac{\hat\mu}{3m+
\sqrt{m^2+4\varphi^2}}\Big)\ .
}
{\bf (9)} $A^\pm$ - $A^\pm$ - $A^3$
\SP{
&\frac{g^2}{192\pi^2m^2(m^2+4\varphi^2)}\Big(
-20m^6-224m^4\varphi^2-224m^2\varphi^4\\ &+(-40m^5+10m^3\varphi^2
+12m\varphi^4
)\sqrt{m^2+4\varphi^2}\\ &+12m^2(13m^4+56m^2\varphi^2+16\varphi^4-
(14m^3-24m\varphi^2)\sqrt{m^2+4\varphi^2})\log\frac{\hat\mu}{\sqrt{m^2+4\varphi^2}}\\ &
+3(-2m^6-15m^4\varphi^2-29m^2\varphi^4-4\varphi^6\\ &+(2m^5+11m^3\varphi^2+11m
\varphi^4)\sqrt{m^2+4\varphi^2})\log\frac{2{\hat\mu}}{-m+\sqrt{m^2+4\varphi^2}}
\\ &+(-3m^6+48m^2\varphi^4+(-3m^5-90m^3\varphi^2-66m\varphi^4)\sqrt{m^2+\varphi^2}
)\log\frac{2{\hat\mu}}{m+\sqrt{m^2+4\varphi^2}}\\ &+(156m^6+672m^4\varphi^2+192m^2
\varphi^4+(168m^5+288m^3\varphi^2)\sqrt{m^2+4\varphi^2})\log\frac{\hat\mu}{2m+\sqrt{
m^2+4\varphi^2}}\\
&+(3m^6+45m^4\varphi^2+135m^2\varphi^4+12\varphi^6\\ &+(-3m^5
+57m^3\varphi^2+33m\varphi^4)\sqrt{m^2+4\varphi^2})\log\frac{2{\hat\mu}}{3m+\sqrt{
m^2+4\varphi^2}}\Big)\ .
}
{\bf (10)} $c^\pm$ - $c^\pm$ - $A^3$
\EQ{
\frac{g^2m^2}{32\pi^2}\log\frac{\hat\mu} m\ .
}
{\bf (11)} $c^3$ - $c^\pm$ - $A^\pm$
\SP{
&-\frac{g^2}{32\pi^2\sqrt{m^2+4\varphi^2}}\Big((m^3+3m\varphi^2-(m^2+\varphi^2)
\sqrt{m^2+4\varphi^2})\log\frac{2{\hat\mu}}{-m+\sqrt{m^2+4\varphi^2}}\\ 
&-(m^3+3m\varphi^2+
(m^2+\varphi^2)\sqrt{m^2+4\varphi^2})\log\frac{2{\hat\mu}}{m+\sqrt{m^2+4\varphi^2}}
\Big)\ .
}
The remaining diagrams are figure-of-eights. The only ones that make a 
non-zero contribution are:\newline
{\bf (12)} $A^\pm$ - $A^3$
\EQ{
\frac{g^2m(m^2+2\varphi^2)}{3\pi^2\sqrt{m^2+4\varphi^2}}\ .
}
{\bf (13)} $A^\pm$ - $A^\pm$
\EQ{
\frac{g^2(m^2+2\varphi^2)^2}{6\pi^2(m^2+4\varphi^2)}\ .
}

Summing up all the graphs, the 2-loop contribution to the effective
potential is
\EQ{
V_\text{2-loop}(\varphi)=\frac{g^2m^2\varphi^2}{8\pi^2(m^2+4\varphi^2)}\ .
\label{res}
}
Notice that all dependence on the dimensional regularization 
scale $\mu$ disappears as one
expects for a finite theory. The other consistency check is that the
potential must vanish when $m=0$, since then the  
theory has $\N=2$ supersymmetry and the effective potential cannot
receive perturbative contributions due to holomorphy. The other
property in its favour is that the potential is smooth as $\varphi\to0$.
The danger is that as $\varphi\to0$, the charged components
of $(\phi^\pm,\psi^\pm)$ becomes massless and have to be included in
the low-energy description. But when the VEV vanishes an
an $SU(2)$ global symmetry is
restored and the potential is then obtained by simply replacing $\varphi$
with the $SU(2)$ invariant $\sqrt{\varphi^a\varphi^a}$.

\section{Discussion}

We have shown that the binding energy for domain walls in $\N=1$
Yang-Mills can be accounted for by the gauge theory that describes
the collective dynamics of the walls. There are a few unresolved
problems. Firstly, we pointed out that the result for the binding
energy from two-loop perturbation theory in the Acharya-Vafa theory
could be subject to stringy corrections. Secondly, we only worked to
two loops in perturbation theory. Clearly it would be hard to go much
beyond this. Since $g^2 \sim \Lambda / N$, higher {\em odd} loop corrections
seem to generate unwanted contributions in even powers in $1/N$ 
to the binding energy. We believe that these contributions vanish 
for the following reason: the binding energy
is a function of $m^2$ and thus higher order corrections must run in powers
of $g^4 / m^2$. If correct, this is sufficient to rule 
out contributions from odd number of loops.

We only considered the case of two walls. With many walls we can make
the following quick point. The binding energy for $k$ walls at leading
order in $1/N$ is
\EQ{
\Delta T_k=
kT_1-T_k=\frac{\pi\Lambda^3}{24 N}(k^3-k)+\cdots\ .
}
In order to understand the $k$ dependence, we write this as
\SP{
\Delta
T_k&=\frac{\pi\Lambda^3}{4N}\Big(\frac{k(k-1)(k-2)}6+\frac{k(k-1)}2\Big)
+\cdots\\
&=
\frac{\pi\Lambda^3}{4N}\Big(\text{\#triples}+\text{\#pairs}\Big)+\cdots\ .
}
So the $k$ dependence is what one expects from the combinatorics of
the pants diagram. Our $U(2)$ result gives the correct coefficient for
the sum over pairs; we shall present the calculation for the triples
elsewhere. 

There are various aspects of the world-volume theory that we postpone
for future
work. It is potentially interesting to investigate the effect of soft
SUSY breaking in the bulk on the world-volume theory.
Na\"\i vely one would expect that the world-volume
fermions will
acquire a mass and hence lead to a non-vanishing one-loop contribution to the
vacuum energy, corresponding to the annulus diagram in string theory. 
Another fascinating aspect of the Acharya-Vafa 
theory is the possibility
of a Seiberg like duality between the $U(k)$ and the $U(N-k)$ world-volume
theories. If the world-volume theory describes $k$ walls it should be
equivalent in some sense 
to the theory of $N-k$ (anti-)walls. 
It would be interesting to see if such a duality could be established.
  
\vspace{1cm}

{\bf Acknowledgments:}
We would like to thank Valya Khoze, Prem Kumar and Mikhail Shifman for discussions. The work of AA is supported by a PPARC advanced fellowship.

\startappendix

\Appendix{Two-Loop Integrals}

In this appendix we describe how to evaluate two-loop integrals of the
form
\EQ{
C(n_1,n_2,n_3)=
\mu^{4\epsilon}
\int\frac{d^dp_1}{(2\pi)^d}\frac{d^dp_2}{(2\pi)^d}\frac{p_1^{2n_1}p_2^{2n_2}
p_3^{2n_3}}
{(p_1^2+m_1^2)(p_2^2+m_2^2)(p_3^2+m_3^2)}\ ,
}
where $d=3-2\epsilon$ and $p_3=p_1+p_2$. The necessary 
techniques can be found
in \cite{Farakos:1994kx} (see also the useful reference
\cite{Kripfganz:1995jx}). Using the tricks 
\EQ{
\frac{p^2}{p^2+m^2}=1-\frac{m^2}{p^2+m^2}\ ,\qquad
\int\frac{d^dp}{(2\pi)^d}\frac{d^dk}{(2\pi)^d}\frac{p^ak^b}{k^2+m^2}=0\ ,
\label{tricks}
}
the latter valid in dimensional regularization,
one can set up the recursion relation
\EQ{
C(n_1,n_2,n_3)=-m_1^2C(n_1-1,n_2,n_3)+(-m_2^2)^{n_2}(-m_3^2)^{n_3}Q_1(n_1-1)\ ,
}
where
\EQ{
Q_1(n)=\mu^{4\epsilon}\int\frac{d^dp_2}{(2\pi)^d}\frac{d^dp_3}
{(2\pi)^d}\frac{p_1^{2n}}{(p_2^2+m_2^2)(p_3^2+m_3^2)}\ .
}
There are similar recursion relations in $p_2$ and $p_3$. By using,
\eqref{tricks} along with
\SP{
R_1(n)&=\mu^{4\epsilon}
\int\frac{d^dp_2}{(2\pi)^d}\frac{d^dp_3}{(2\pi)^d}\frac{(p_2\cdot
  p_3)^n}{(p_2^2+m_2^2)(p_3^2+m_3^2)}\\ &=\begin{cases}
\frac1{d+n-2}(-m_2^2)^{n/2}
(-m_3^2)^{n/2}I(m_2,m_3) & n\text{ even}\\ 0 & n\text{ odd}\end{cases}\ ,
}
where 
\EQ{
I(m_2,m_3)=\mu^{4\epsilon}
\int\frac{d^dp_2}{(2\pi)^d}\frac{d^dp_3}{(2\pi)^d}\frac{1}
{(p_2^2+m_2^2)(p_3^2+m_3^2)}=\frac{m_2m_3}{16\pi^2}\ .
}
one has
\EQ{
Q_1(n)=\sum_{a_1,a_2,a_3\atop a_1+a_2+a_3=n}\frac{n!}{a_1!a_2!a_3!}
2^{a_3}(-m_2^2)^{a_2}(-m_3^2)^{a_3}R_1(a_3)\ .
\label{fur}
}
The recursion relations along with the basic result
\SP{
C(0,0,0)&=
\mu^{4\epsilon}\int\frac{d^dp_1}{(2\pi)^d}\frac{d^dp_2}{(2\pi)^d}\frac{1}
{(p_1^2+m_1^2)(p_2^2+m_2^2)(p_3^2+m_3^2)}\\ &=\frac1{16\pi^2}\Big(
\frac1{4\epsilon}-\frac\gamma2+\frac12\log4\pi+\frac12+\log\frac\mu{m_1
+m_2+m_3}\Big)\ ,
}
and \eqref{fur} 
can be used to evaluate all the integrals encountered at two loops.

\end{document}